\documentclass{article}
\usepackage{spconf,amsmath,graphicx,amssymb,algorithm,algpseudocode,hyperref,siunitx,etoolbox}

\def\L{{\cal L}}

\title{Intelli-z: toward intelligible zero-shot TTS}
\name{Sunghee Jung, Won Jang, Jaesam Yoon, Bongwan Kim}
\address{Kakao Brain, Seongnam, Republic of Korea}
\begin{document}
%\ninept
%
\maketitle
\begin{abstract}
% Although numerous recent studies have suggested new paradigms for zero-shot TTS with large-scale, real-world data, studies that focus on the intelligibility of zero-shot TTS are relatively scarce. Zero-shot TTS demands additional efforts to ensure clear pronunciation and speech quality owing to its inherent requirement of replacing a core parameter(speaker embedding or acoustic prompt) with a new one at the inference stage. In this study, we propose a zero-shot TTS focused on intelligibility and dub this model Intelli-Z. Intelli-Z learns speaker embeddings from prototype embeddings and is trained with meta-learning assumption to include query set. Also, it selectively aggregates speaker embedding in temporal dimension to minimize the interference of text content of reference speech at inference stage. We substantiate the effectiveness of proposed methods with an ablation study. Mean opinion score(MOS) is increased for unseen speakers by 9\% compared to baseline when the first two methods are applied and it is further improved by 16\% when the selective temporal aggregation is applied. 
% In this study, we propose two loss functions and an aperiodicity mask to enhance the intelligibility of zero-shot TTS. This proposed model is named as Intelli-Z. To validate our approach, we conducted a subjective listening test and successfully demonstrated the effectiveness of our proposed method.
Although numerous recent studies have suggested new frameworks for zero-shot TTS using large-scale, real-world data, studies that focus on the intelligibility of zero-shot TTS are relatively scarce. Zero-shot TTS demands additional efforts to ensure clear pronunciation and speech quality due to its inherent requirement of replacing a core parameter (speaker embedding or acoustic prompt) with a new one at the inference stage.
In this study, we propose a zero-shot TTS model focused on intelligibility, which we refer to as Intelli-Z. Intelli-Z learns speaker embeddings by using multi-speaker TTS as its teacher and is trained with a cycle-consistency loss to include mismatched text-speech pairs for training. Additionally, it selectively aggregates speaker embeddings along the temporal dimension to minimize the interference of the text content of reference speech at the inference stage.
We substantiate the effectiveness of the proposed methods with an ablation study. The Mean Opinion Score (MOS) increases by 9\% for unseen speakers when the first two methods are applied, and it further improves by 16\% when selective temporal aggregation is applied.
\end{abstract}
\section{Introduction}
Zero-shot text-to-speech (TTS) is a system capable of synthesizing speech in an unfamiliar voice, making the disentanglement of text and speaker information a central challenge. In the context of multi-speaker TTS, a speaker embedding lookup table is typically employed, where each speaker embedding is learned inductively by marginalizing characteristics dependent on the given text input. This approach facilitates disentangling text content from speaker identity.

However, in zero-shot TTS, the process involves extracting speaker embeddings from varying sentences during the training stage using a speaker encoder. Therefore, the separation of speaker and text information becomes difficult. Several studies have addressed the zero-shot TTS problem, including notable studies, such as the proposal of transfer learning from speaker verification by Jia et al\cite{jia2018transfer}. The speaker verification model can leverage larger-scale speech data from a more diverse set of speakers since it can tolerate noise to some extent, unlike generation tasks. By incorporating the speaker encoder of the speaker verification model into zero-shot TTS, Jia et al. significantly improved the quality of zero-shot TTS.
Additionally, Meta-StyleSpeech\cite{meta} introduced the use of adversarial loss, enabling the training of zero-shot TTS with speaker-text pairs that cannot be directly trained using reconstruction loss. Numerous other works have contributed to the field of zero-shot TTS\cite{nautilus,content, yourtts}.

Influenced by the recent development of large-scale language models, there has been a surge in research focusing on zero-shot TTS. Recently, novel approaches have emerged, aiming to leverage pretrained neural audio codecs\cite{encodec} to convert audio signals into discretized tokens and apply next-token prediction objectives for training zero-shot TTS models\cite{audiolm,valle,speartts,speechx}. Since both the discretization of audio signals into tokens and next-token prediction are unsupervised learning objectives, these methods can effectively utilize vast amounts of real-world data.

Nevertheless, regardless of the selected paradigm, the poor intelligibility of zero-shot TTS systems remains a challenge. This is primarily due to the fundamental problem of substituting a core parameter of the model with an unseen one, such as a speaker embedding or acoustic prompt, as seen in recent works. Surprisingly, studies dedicated to addressing the intelligibility of zero-shot TTS systems are currently lacking. Herein, we focused on enhancing the intelligibility of zero-shot TTS; accordingly, the proposed model was named Intelli-Z. 

\begin{figure*}[t]
  \centering
  \includegraphics[width=\linewidth]{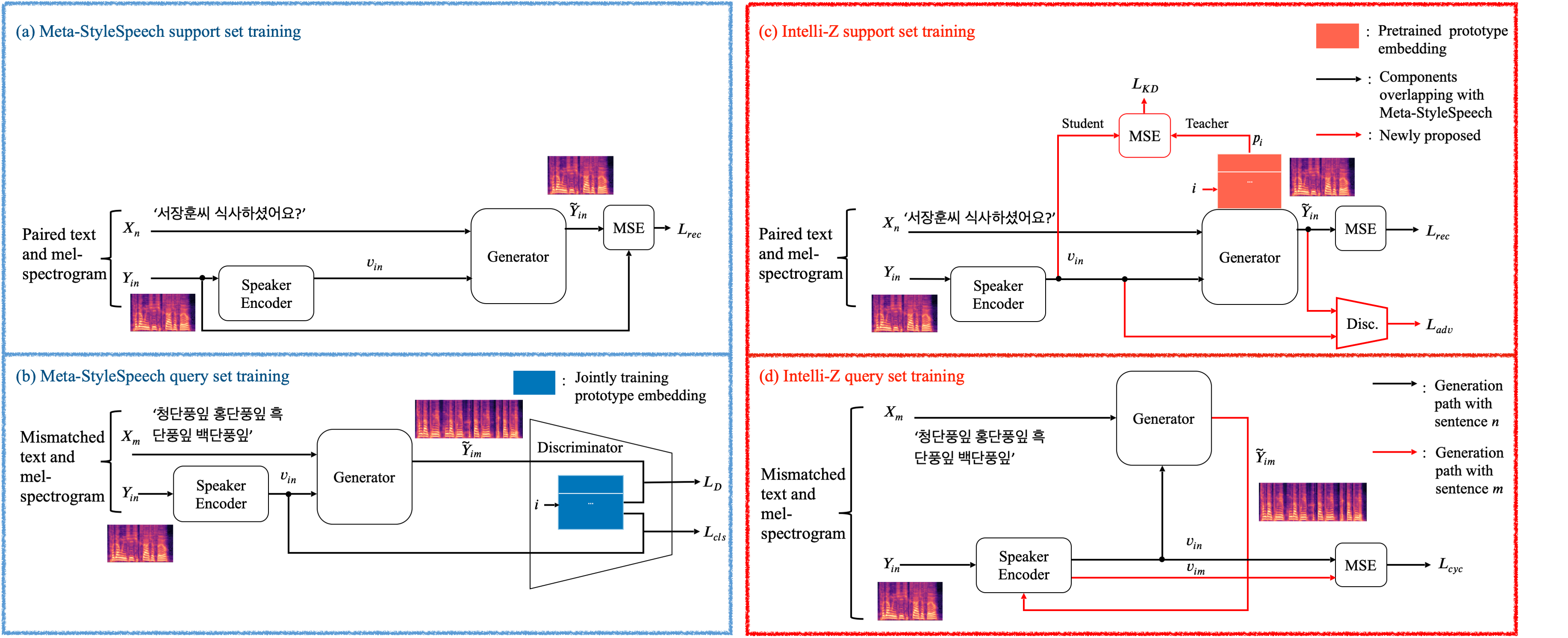}
  \caption{(a): Support set training scheme of Meta-StyleSpeech. Text encoder and text discriminator are omitted in the figure for simplicity. The variance adaptor and decoder blocks are inside the generator, as in FastSpeech2. (b): Query set training scheme of Meta-StyleSpeech. (c): Support set training scheme of Intelli-Z. Text encoder is omitted for simplicity. (d): Query set training scheme of Intelli-Z.}
  \label{fig:one_fig}
\end{figure*}
In this paper, we present three main contributions:
\begin{itemize}
    \item Knowledge Distillation: We present the distillation of knowledge from multi-speaker TTS into the realm of zero-shot TTS. This transfer of knowledge enhances the capabilities of zero-shot TTS systems.
    \item Cycle-Consistency for Robustness: To ensure robustness across inference conditions, we introduce the use of cycle-consistency. This approach enhances the stability and reliability of zero-shot TTS models.
    \item Selective Voice Frame Aggregation: We use only voiced frames when temporally aggregating the speaker encoder output. This selective approach enhances the effectiveness of zero-shot TTS systems in capturing essential speaker characteristics.  
\end{itemize}
\section{Meta-StyleSpeech}
% \begin{figure}[t]
%   \centering
%   \includegraphics[width=\linewidth]{meta_0308.png}
%   \caption{Diagram of Meta-StyleSpeech. Text encoder and text discriminator are omitted in the figure for simplicity. Inside the generator are variance adaptor and decoder blocks as in FastSpeech2. Upper figure: support set training scheme. Lower figure: query set training scheme}
%   \label{fig:metastylespeech}
% \end{figure}
\subsection{Support set and query set}
\label{subsub}
Meta-StyleSpeech introduces meta-learning concepts to the domain of zero-shot TTS. A defining characteristic of meta-learning is the division of the training dataset into two distinct components: support set and query set. The support set comprises data that are accessible during the training phase, such as paired speech and script samples. In contrast, the query set is employed to simulate the inference environment, which may differ from the training environment\cite{episodic}.
In the context of Meta-StyleSpeech, this discrepancy manifests as a situation where the input text is unrelated to the content of the reference speech. The support set is denoted as $(X_{n}, Y_{in})$, with $X_n$ representing the $n$-th sentence and $Y_{in}$ representing the $n$-th speech sample of speaker $i$. Conversely, the query set is represented as $(X_m, Y_{in})$, with $X_m$ denoting the $m$-th sentence, where $n \neq m$.
\subsection{Adversarial loss for training query set}
Fig. \ref{fig:one_fig}. (a) and (b) illustrate the training process in Meta-StyleSpeech for the support and query sets, respectively. The support set is trained using a reconstruction loss ($L_{rec}$) based on the ground truth mel-spectrogram $Y_{in}$. However, the absence of a ground truth ($Y_{im}$) for the query set necessitates some special approach. Meta-StyleSpeech employs a discriminator that learns prototype embeddings ($p_i$) to optimize both cross-entropy loss ($L_{cls}$) aligning speaker encoder outputs with the discriminator and CGAN projection discriminator loss ($L_D$) to train the mel spectrogram of the query set $\tilde{Y}_{im}$\cite{meta,miyato}.
%%%%%%%%%%%여기서부터 다시 볼 것. 09.10.일.
\section{Proposed idea}
\subsection{Knowledge distillation from multi-speaker TTS}
The primary difference between Meta-StyleSpeech and our proposed model lies in the definition of the prototype embedding. First, we conduct pretraining for multi-speaker TTS before moving on to zero-shot TTS. During the zero-shot TTS model training phase, the speaker encoder learns from the speaker embeddings sourced from the lookup table of the pre-trained multi-speaker TTS model (referred to as prototype embedding) as its ground truth. Once trained, the speaker encoder has the capability to generate new speaker embeddings for previously unseen speakers during inference. These newly generated embeddings emulate the behavior of the prototype embeddings and produce intelligible speech.
$v_{in}$ represents the speaker embedding extracted using the speaker encoder $E_s$ with the mel-spectrogram $Y_{in}$ as input. Here, $i$ denotes the index of the speaker, and $n$ is the index of the sentence. The knowledge distillation loss, denoted as $L_{KD}$, is computed as the $L2$ norm of the difference between the prototype embedding $p_i$ and the speaker embedding $v_{in}$. The prototype embedding $p_i$ is obtained from the lookup table of the pretrained multi-speaker TTS model.
% \begin{equation}
% \begin{split}
\begin{gather}
{v_{in}}=E_s(Y_{in}),\\
L_{KD}=\mathop{\mathbb{E}}_{i\in I}\mathop{\mathbb{E}}_{n\in N_i}\lVert{}p_i-v_{in}\rVert_2
\label{eq:kd}
\end{gather}
% \end{split}
% \end{equation}
\begin{equation}
\begin{split}
    L_{disc}=\mathop{\mathbb{E}}_{i\in I}\mathop{\mathbb{E}}_{n\in N_i}[\{D({Y_{in}},v_{in})-1\}^2+\\
    \{D(G(X_{n},v_{in}),v_{in})\}^2].
\label{eq:adv}
\end{split}
\end{equation}
In Figure \ref{fig:one_fig}.(c), we employ $L_{KD}$ in conjunction with the reconstruction loss ($L_{rec}$) and GAN loss ($L_{adv}$) for training the support set. The equation for the GAN loss is presented in Eq. \ref{eq:adv}. The generator is trained with the adversarial counterpart of Eq. \ref{eq:adv}. In this context, we use a CGAN projection discriminator, which utilizes the speaker embedding as a condition to evaluate the mel-spectrogram. The entities denoted as $G$ and $D$ represent the generator and discriminator, respectively. 
% The discriminator is trained to predict 1 when presented with the ground truth mel-spectrogram $Y_{in}$ as input and the speaker embedding $v_{in}$ as a condition. Conversely, the discriminator is trained to predict 0 when given the generated support set mel-spectrogram $G(X_n,v_{in})$ as input and $v_{in}$ as a condition.

%%%%%%%%%%%%%%%%%%%%%%%%%%%%%%%%%%%%%%%%%%%%%%%%%%%%%%%%%%%%%%%%%%%%%%%%%%%%%%%%%%%%%%%%%5
\subsection{Cycle-consistency loss}
The query set is denoted as $(X_m, Y_{in})$, as mentioned in Section \ref{subsub}. The training of query set is illustrated in Fig. \ref{fig:one_fig} (d). We define the cycle-consistency as in Eq. \ref{eq:cyc}. Initially, $v_{in}$ is extracted from the speaker encoder using $Y_{in}$ as input, following which the mel-spectrogram $\tilde{Y}_{im}$ is generated with query text $X_m$ and speaker embedding $v_{in}$ as inputs. The text encoder is omitted in Fig. \ref{fig:one_fig}. Subsequently, speaker embedding $v_{im}$ is extracted from this $\tilde{Y}_{im}$, representing the speaker $i$ and text $m$. We hypothesize that a speaker embedding that is robust to changes in the text embedding will yield a $v_{im}$ similar to $v_{in}$.
$L_{cyc}$ is calculated by averaging the $L2$ norm of the difference between $v_{im}$ and $v_{in}$ for all text pairs $(n,m)$ where $n \neq m$ across all speakers $i$.
\begin{gather}
\tilde{Y}_{im}=G(X_{m},{v_{in}}),
\\{v_{im}}=E_s(\tilde{Y}_{im}),
\\L_{cyc}= \mathop{\mathbb{E}}_{i\in I}\mathop{\mathbb{E}}_{n,m\in N_i}\lVert{}v_{im}-{v_{in}}\rVert_2
\label{eq:cyc}
\end{gather}
The final equation for the generator is as in Eq.\ref{eq:final}.
\begin{equation}
    L_{gen}=\lambda_{KD}L_{KD}+\lambda_{cyc}L_{cyc}+\lambda_{adv}L_{adv}+L_{rec}
\label{eq:final}
\end{equation}
\subsection{Selective Voice Frame Aggregation}
It is essential to temporarily aggregate the output of the speaker encoder. When performing this, we employed a multi-head self-attention mechanism, followed by averaging the weighted values over time. 

During internal listening tests, we observed that the model tended to confuse unvoiced phonemes from the reference with speaker identity. Additionally, we made an assumption that speaker identity is more discernible in voiced frames than that in unvoiced frames. Building on this insight, we were motivated to include only voiced frames while excluding unvoiced frames during the aggregation process. To achieve this, we introduced an aperiodicity mask and used it in the computation of the score for the multi-head attention.
$periodicity_t$ in Eq. \ref{eq:periodicity} is defined as described in \cite{d4c}. The methodology for acquiring this information is elaborated in Section \ref{sec:exp}.
\begin{equation}
{Mask}_t=
\begin{cases}
  0, & \text{if}\ {periodicity}_t<1.0 \\
  1, & \text{otherwise}
\end{cases}
\label{eq:periodicity}
\end{equation}
\begin{equation}
    Q, K, V = W^{Q}s_{i},W^{K}s_{i},W^{V}s_{i},
\end{equation}
\begin{equation}
    Scores = \frac{QK^T}{\sqrt{d}}
\end{equation}
\begin{equation}
    {Scores}_t=
\begin{cases}
  -\infty,      & \text{if}\ {Mask}_t = 0\\
  {Scores}_t, & \text{otherwise}
\end{cases}
\end{equation}
\begin{equation}
    \hat{s}_{i}=Softmax(Scores)V
\label{eq:softmax}
\end{equation}
\begin{equation}
v_i=\frac{1}{T} \sum_{t=1}^{T}\hat{s}_t.
\end{equation}\\
Here, $s_i \in \mathbb{R}^{d \times T}$ represents the speaker embedding sequence before temporal aggregation. The variable $i$ serves as the index for the speaker, while $d$ denotes the dimension of the speaker embedding, and $T$ indicates the length of the reference frame. Furthermore, we have $W^{Q}, W^{K}, W^{V} \in \mathbb{R}^{h \times d}$, where $h$ represents the number of heads.
As the $Scores$ are set to $-\infty$ for aperiodic frames before the softmax operation in Eq. \ref{eq:softmax}, the scores for these frames will subsequently converge to zero after applying the softmax. Consequently, only the frames containing voiced signals will retain non-zero values in $\hat{s}_{i}$. $v_i$ represents the speaker embedding, which is referred to as $v_{in}$ in other sections of this paper. The subscript $n$ is omitted since the reference text index is irrelevant in this section.
\section{Experiments and results}
\label{sec:exp}
\subsection{Dataset}
We utilized a publicly available Korean multi-speaker corpus \cite{25}. This corpus comprised three subsets, with 2,400 / 4,800 / 12,000 sentences per person, respectively. The subset containing 2,400 sentences per person was employed for training, involving 1,872 speakers. The remaining subsets were reserved for evaluation.
% `Unseen speakers' are selected from remaining subsets of the same corpus.
\subsection{Model details}
The dataset was resampled at 24 kHz. We used a window size of 1024, FFT size of 1024, hop size of 256, and 100 mel-spectrogram bins. We employed the conformer-based JDI-T \cite{23,21} and UnivNet \cite{jang21_interspeech} for the TTS component and vocoder, respectively. To generate the aperiodicity mask, we utilized the open-source software `pyworld'\footnote{https://github.com/JeremyCCHsu/Python-Wrapper-for-World-Vocoder}. The speaker encoder structure of our model matches that of Meta-StyleSpeech. All hyperparameter details are consistent with those of the aforementioned papers. After a series of experiments, we have determined the loss weights as follows: $\lambda_{KD}$ = 0.5, $\lambda_{cyc}$ = 0.5, and $\lambda_{adv}$ = 0.1.

% Dataset was resampled at 24 kHz. 1024 window size, 1024 FFT size, 256 hop size, and 100 mel-spectrogram bins are used. We used conformer-based JDI-T \cite{23,21} for TTS and UnivNet \cite{jang21_interspeech} for vocoder. For generating the aperiodicity mask, the open source software `pyworld'\footnote{https://github.com/JeremyCCHsu/Python-Wrapper-for-World-Vocoder} was used.
% % For the implemenetation of Meta-StyleSpeech, we also used conformer-based JDI-T instead of FastSpeech2\cite{24} for fair comparison. 
% The speaker encoder structure of our model is the same as Meta-StyleSpeech. All hyperparameter details are equal to that of aforementioned papers. The loss weights $\lambda_{KD}$, $\lambda_{cyc}$, $\lambda_{adv}$ are 0.5, 0.5 and 0.1.
\subsection{Subjective evaluation}
Listeners were requested to assess the speech quality and similarity on a scale of 1 to 5, referred to as MOS and SMOS, respectively. A total of 14 participants contributed to the MOS ratings, while 15 individuals participated in the SMOS evaluations.
When comparing Table \ref{tab:seen} and Table \ref{tab:unseen}, it was consistently observed across all three models that both MOS and SMOS scores dropped for unseen speakers.
Notably, Meta-StyleSpeech consistently underperformed our proposed models for both seen and unseen speakers in terms of MOS as well as SMOS.
As discussed in \cite{investigating}, the use of discriminative losses for learning speaker embedding can lead to distinct decision boundaries but may hinder generative capabilities. Consequently, our proposed prototype training method demonstrated greater stability across a diverse range of speakers when contrasted with Meta-StyleSpeech.
In Table \ref{tab:seen} and \ref{tab:unseen}, the label `Intelli-Z w/o ap' represents Intelli-Z without the aperiodicity mask. It is noteworthy that the application of the aperiodicity mask significantly improved both the MOS and SMOS scores for Intelli-Z. 
Although our aim was to increase intelligibility, it was observed that improved intelligibility also led to an increased perceived similarity.
Further details on sample outputs are accessible online\footnote{https://jsh-tts.tistory.com/entry/icassp2024}.
\begin{table}[h]
\sisetup{
  table-align-uncertainty=true,
  separate-uncertainty=true,
}
%% local redefinitions
\renewrobustcmd{\bfseries}{\fontseries{b}\selectfont}
\renewrobustcmd{\boldmath}{}
\centering
\caption{Subjective evaluation for seen speakers.}
\label{tab:seen}
\begin{tabular}{l|cc}
\noalign{\smallskip}\noalign{\smallskip}\hline\hline
                                    & MOS              & SMOS                 \\ \hline
Meta-StyleSpeech                    &3.25  $\pm$ 0.10  & 3.62  $\pm$ 0.08      \\
Intelli-Z w/o ap                    &3.58  $\pm$ 0.10  & 3.80  $\pm$ 0.08      \\ 
Intelli-Z                           &\textbf{4.01  $\pm$ 0.08}  & \textbf{3.97  $\pm$ 0.07}      \\
Ground truth                        & 4.88 $\pm$ 0.04  & 4.23  $\pm$ 0.10      \\
\hline\hline
\end{tabular}
\end{table}
\begin{table}[h]
\sisetup{
  table-align-uncertainty=true,
  separate-uncertainty=true,
}
%% local redefinitions
\renewrobustcmd{\bfseries}{\fontseries{b}\selectfont}
\renewrobustcmd{\boldmath}{}
\centering
\caption{Subjective evaluation for unseen speakers.}
\label{tab:unseen}
\begin{tabular}{l|cc}
\noalign{\smallskip}\noalign{\smallskip}\hline\hline
                                  &      MOS           & SMOS \\ \hline
                                  
Meta-StyleSpeech                  &2.86 $\pm$ 0.11     &3.45 $\pm$ 0.09        \\
Intelli-Z w/o ap                  &3.13 $\pm$ 0.10     &3.48 $\pm$ 0.09        \\ 
Intelli-Z                         &\textbf{3.65 $\pm$ 0.11}     &\textbf{3.80 $\pm$ 0.08}       \\
Ground truth                      &4.65 $\pm$ 0.06     &4.09 $\pm$ 0.09       \\
\hline\hline
\end{tabular}
\end{table}
\subsection{Qualitative evaluation of aperiodicity masks}
\begin{figure}
  \centering
  \includegraphics[width=\linewidth]{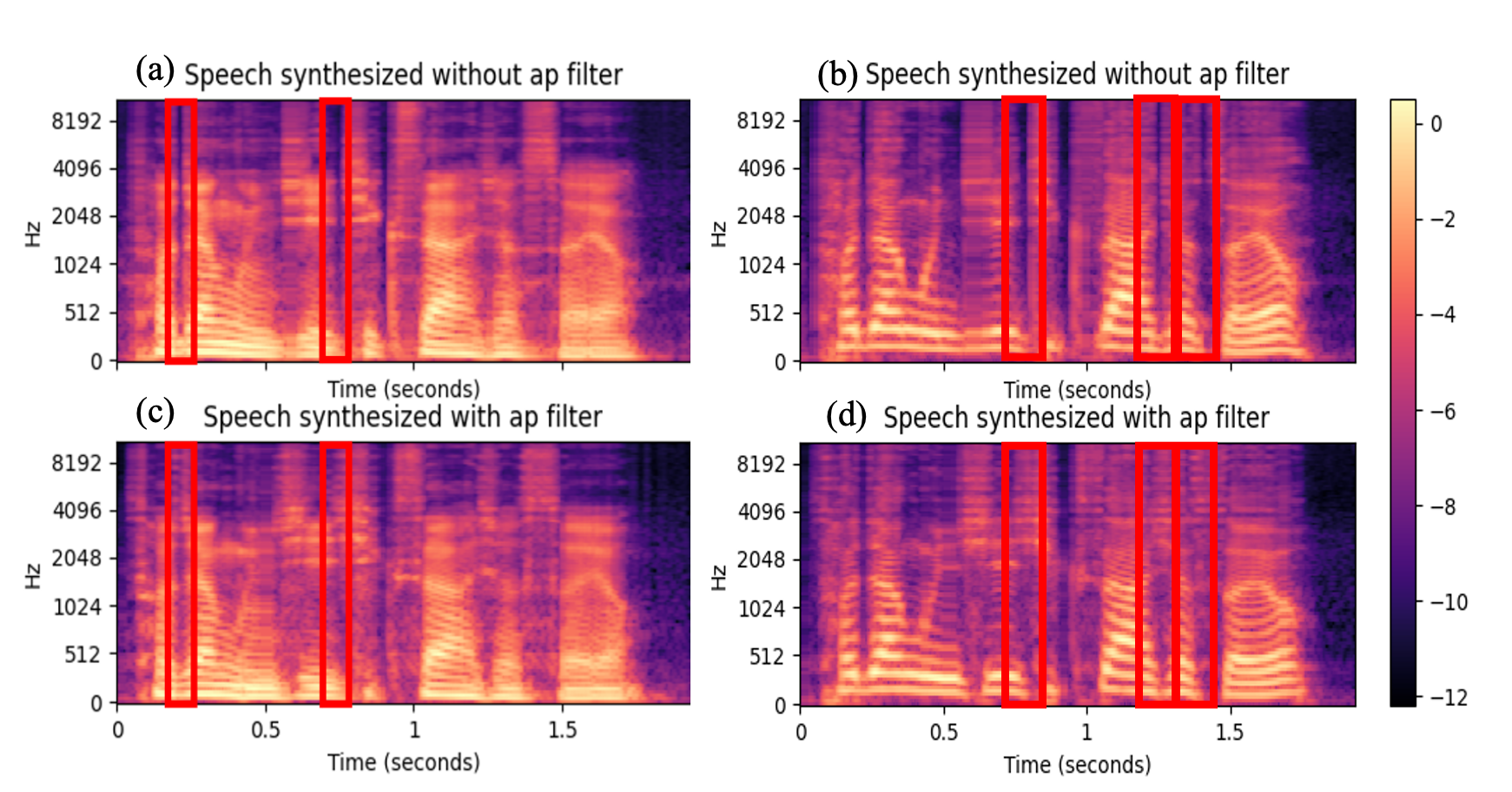}
  \caption{(a), (b): Mel-spectrograms synthesized without applying the aperiodicity mask. (c), (d): Mel-spectrograms synthesized with the application of the aperiodicity mask. The black lines observed in the red boxes in (a) and (b) result in lisping sounds, and /s/ sounds become /th/ or /d/. This problem is resolved in (c) and (d).}
  \label{fig:ap_result}
\end{figure}

% We visualized the effectiveness of the aperiodicity mask.
In this subsection, we further analyze the effect of aperiodicity mask qualitatively.
In Fig. \ref{fig:ap_result}, (a) and (b) represent the mel-spectrograms of speech synthesized using a model trained and inferred without the aperiodicity masks, while (c) and (d) showcase the mel-spectrograms of the synthesized speech with the aperiodicity masks applied.
In the boxed regions of Fig. \ref{fig:ap_result}. (a) and (b), discernible black lines are evident. Conversely, these black lines in the higher frequencies are replaced with noisy frames in (c) and (d). These black lines are responsible for generating a lisping sound in voice cloning scenarios\footnote{Samples are included in the demo sample webpage}. We attribute the improvement in the unvoiced region to the fact that the unvoiced characteristics are trained to be solely controlled by the text and remain unaffected by the reference speech. Consequently, the aperiodicity mask contributes to better disentanglement between the text embedding and speaker identity. This observation was consistently made across various tested speakers.
\section{Conclusion}
This study aimed to improve the intelligibility of zero-shot TTS using three suggested methods. First, we proposed leveraging multi-speaker TTS speaker embeddings as prototypes for zero-shot TTS speaker embedding training, promoting its coordination with text embedding. Second, we introduced cycle-consistency loss to address training and inference disparities, mitigating the impact of textual content in reference speech. Finally, we suggested selective temporal aggregation using aperiodicity features and this significatly improved intelligibility by eliminating lisping sounds. We proved the effectiveness of our proposed methods in an ablation study. Suggested losses enhanced the MOS by 9\% compared to the baseline and adding aperidocity mask further improved the MOS by 16\% on unseen speakers. Also, we qualitatively analyzed the effect of aperiodicity mask on removing lisping sounds by visualizing and comparing the mel-spectrograms.
% Our approach was validated through a subjective listening test. the results present clear visualizations illustrating the benefits of selective aggregation, collectively advancing zero-shot TTS.
% Below is an example of how to insert images. Delete the ``\vspace'' line,
% uncomment the preceding line ``\centerline...'' and replace ``imageX.ps''
% with a suitable PostScript file name.
% -------------------------------------------------------------------------

%\vfill
%\pagebreak

\bibliographystyle{IEEEbib}
\bibliography{main}

\end{document}